\documentclass[leqno,12pt]{amsart}
\usepackage{amsfonts,latexsym}
\usepackage{amsmath}
\usepackage{amscd}
\usepackage{float,amsmath,amssymb,mathrsfs,bm,multirow,graphics}
\usepackage[dvips]{graphicx}

\newcommand{\nequation}{\setcounter{equation}{0}}

\newcommand{\R}{{\Bbb R}}

\newcommand{\C}{{\Bbb C}}

\newcommand{\proofbegin}{\noindent{\it Proof.\,\,}}
\newcommand{\proofend}{\hfill$\Box$\bigskip}

\newtheorem{theorem}{Theorem}[section]
\newtheorem{proposition}[theorem]{Proposition}
\newtheorem{lemma}[theorem]{Lemma}

\input epsf
\title[Dressing for a novel integrable generalization]{\sc Dressing for a novel integrable generalization of the nonlinear Schr\"odinger equation}
\author{Jonatan Lenells}
\address{Institut f\"ur Angewandte Mathematik, Leibniz Universit\"at Hannover \\
Welfengarten 1, 30167 Hannover, Germany}
\email{lenells@ifam.uni-hannover.de}

\begin{document}
\begin{abstract} 

\noindent
We implement the dressing method for a novel integrable generalization of the nonlinear Schr\"odinger equation. As an application, explicit formulas for the $N$-soliton solutions are derived. As a by-product of the analysis, we find a simplification of the formulas for the $N$-solitons of the derivative nonlinear Schr\"odinger equation given by Huang and Chen.
\end{abstract}

\maketitle



\section{Introduction} \nequation
We consider the following integrable generalization of the nonlinear Schr\"odinger (NLS) equation, which was first derived in \cite{F} by means of bi-Hamiltonian methods:
\begin{equation}\label{GNLS}  
  iu_t - \nu u_{tx} + \gamma u_{xx} + \sigma |u|^2(u  + i \nu u_x) = 0, \qquad \sigma = \pm 1,
\end{equation}
where $\gamma$ and $\nu$ are nonzero real parameters and $u(x,t)$ is a complex-valued function.  Equation (\ref{GNLS}) reduces to the NLS equation when $\nu = 0$. However, the limit $\nu \to 0$ is in many respects singular, and the analysis of (\ref{GNLS}) is rather different from that of NLS. Equation (\ref{GNLS}) appears as a model for nonlinear pulse propagation in optical fibers provided that one retains certain terms of the next asymptotic order beyond those necessary for the NLS equation \cite{F, Lfiber}. From a bi-Hamiltonian point of view, equation (\ref{GNLS}) is related to NLS in the same way that the celebrated Camassa-Holm equation \cite{C-H} is related to KdV \cite{F}. Being integrable, (\ref{GNLS}) admits a Lax pair formulation and the initial-value problem on the line can be analyzed by means of the inverse scattering transform (IST) \cite{LF}. 
The spectral analysis of (\ref{GNLS}) is closely related to that of the derivative NLS equation. In fact, equation (\ref{GNLS}) is related by a gauge transformation to the first negative member of the integrable hierarchy associated with the derivative NLS equation \cite{Lfiber}.
The initial-boundary value problem for equation (\ref{GNLS}) on the half-line was studied in \cite{LF2}.

Here, we implement the dressing method for equation (\ref{GNLS}) and derive, as an application of the general approach, explicit formulas for the $N$-soliton solutions. 
The dressing method is a technique which generates new solutions of an integrable equation from an already known seed solution cf. \cite{NMPZ}. In our implementation of the dressing method to equation (\ref{GNLS}), we start with a seed solution $u_0$ together with a corresponding eigenfunction $\psi_0$ of the Lax pair. A new eigenfunction $\psi = G \psi_0$ is constructed by multiplying $\psi_0$ by a $2\times2$-matrix $G$ which contains an even number of poles in the spectral parameter. It is shown that $\psi$ satisfies the same Lax pair as $\psi_0$ but with the potential $u_0$ replaced with a new potential $u(x,t)$. The compatibility of the Lax pair equations then implies that $u(x,t)$ also is a solution of (\ref{GNLS}). 

An expression for the one-soliton solution of (\ref{GNLS}) was found in \cite{LF} by solving directly the Riemann-Hilbert problem associated with the inverse problem in the presence of two poles and no jump (each soliton contributes {\it two} poles to the Riemann-Hilbert problem because the poles are of necessity symmetrically distributed with respect to the origin). Applying the dressing approach to the particular seed solution $u_0 = 0$ we recover the one-soliton solution and also find an explicit expression for the $N$-soliton solution for any $N$.

A convenient Lax pair for (\ref{GNLS}) is introduced in Section \ref{laxsec}, while the implementation of the dressing method is described in detail in Section \ref{dressingsec}.
In Section \ref{solitonsec} we derive explicit expressions for the $N$-solitons.
Finally, in Section \ref{DNLSsolitonsec}, motivated by the analysis of (\ref{GNLS}), we find a simplification of the formula for the $N$-soliton solution given by Huang and Chen \cite{HC} of the derivative nonlinear Schr\"odinger (DNLS) equation
\begin{equation}\label{DNLS}
  i q_t + q_{xx} + i\left(|q|^2 q\right)_x = 0, \qquad x \in \R, \quad t \geq 0.
\end{equation}
The connection between equations (\ref{GNLS}) and (\ref{DNLS}) arises because the $x$-parts of the corresponding Lax pairs are identical upon identification of $u_x$ with $q$.

\section{A Lax pair}\label{laxsec}\nequation
It was noted in \cite{Lfiber} that the nonzero values of the parameters $\gamma, \nu, \sigma$ in (\ref{GNLS}) can be arbitrarily assigned via a change of variables. We will therefore, for simplicity, henceforth assume that $\gamma = \nu = \sigma = 1$. Then the transformation $u \to e^{ix} u$ converts (\ref{GNLS}) into the equation
\begin{equation}\label{GNLSgauge}  
  u_{tx} + u - 2i u_x - u_{xx} - i |u|^2u_x = 0,
\end{equation}
which admits the following Lax pair \cite{LF}:
\begin{equation}\label{psilax}
\begin{cases}
	& \psi_x + i\frac{1}{\zeta^2} \sigma_3 \psi = \frac{1}{\zeta} U_x \psi,	\\
	& \psi_t + i\left(\frac{1}{\zeta} - \frac{\zeta}{2}\right)^2 \sigma_3 \psi = \left(\frac{1}{\zeta} U_x - \frac{i}{2}\sigma_3U^2 +\frac{i\zeta}{2}\sigma_3U\right)\psi,
\end{cases}
\end{equation}
where $\psi(x,t,\zeta)$ is a $2\times2$-matrix valued eigenfunction, $\zeta \in \hat{\C} = \C \cup \{\infty\}$ is a spectral parameter, and
\begin{equation}\label{Udef}  
  U(x,t) =\begin{pmatrix} 0	& u(x,t)	\\
-\bar{u}(x,t)	&	0 \end{pmatrix}, \qquad \sigma_3 = \begin{pmatrix} 1 & 0 \\ 0 & -1 \end{pmatrix}.
\end{equation}
We will assume that $\psi$ obeys the symmetries
\begin{align}\label{psisymmetry}  
  \sigma_3\psi(x,t,-\zeta)\sigma_3 = \psi(x,t,\zeta),\qquad  \psi^{-1}(x,t,\zeta) = \psi^\dagger(x,t,\bar{\zeta}).	
\end{align}

A solution $\psi$ of (\ref{psilax}) satisfying (\ref{psisymmetry}) can be constructed as follows: 
Assuming that $u$ has sufficient smoothness and decay, there exists a solution of (\ref{psilax}) which tends to $\exp(-i\theta \sigma_3)$ at infinity, where
\begin{equation}\label{thetadef}
   \theta := \theta(x,t,\zeta) = \frac{1}{\zeta^2}x + \left(\frac{1}{\zeta} - \frac{\zeta}{2}\right)^2t.
\end{equation}
This suggests the transformation $\psi = \Psi e^{-i\theta \sigma_3}$ and then $\Psi$ solves
\begin{equation}\label{Psilax}  
\begin{cases}
	& \Psi_x + i\frac{1}{\zeta^2}[\sigma_3, \Psi] = \frac{1}{\zeta}U_x \Psi,	\\
	& \Psi_t + i\left(\frac{1}{\zeta} - \frac{\zeta}{2}\right)^2 [\sigma_3, \Psi] = \left(\frac{1}{\zeta} U_x - \frac{i}{2}\sigma_3U^2 +\frac{i\zeta}{2}\sigma_3U\right)\Psi.
\end{cases}
\end{equation}
We define two solutions $\Psi_1$ and $\Psi_2$ of (\ref{Psilax}) via the linear Volterra integral equations
\begin{subequations}\label{Psi12def}
\begin{align} \label{Psi1def}
  \Psi_1(x, t, \zeta) = I + \int_{-\infty}^x e^{i \zeta^{-2} (x' - x) \hat{\sigma}_3} \zeta^{-1}(U_x \Psi_1)(x', t, \zeta)dx',
 	\\ \label{Psi2def}
  \Psi_2(x, t, \zeta) = I - \int_x^{\infty} e^{i \zeta^{-2} (x' - x) \hat{\sigma}_3} \zeta^{-1}(U_x \Psi_2)(x', t, \zeta)dx',
\end{align}
\end{subequations}
where $\hat{\sigma}_3$ acts on a $2\times 2$ matrix $A$ by $\hat{\sigma}_3A = [\sigma_3, A]$.
The second columns of these equations involves $\exp[2i\zeta^{-2}(x' - x)]$. It follows that the second column vectors of $\Psi_1$ and $\Psi_2$ are well-defined and analytic for $\zeta \in \hat{\C}$ such that $\zeta^2$ lies in the upper and lower half-planes, respectively. Moreover, these column vectors have continuous extensions to $\zeta \in \R \cup i\R \setminus \{0, \infty\}$. Similar remarks apply to the first column vectors. 

We define a solutionÊ $\psi$ of (\ref{psilax}) by
\begin{equation}\label{psiPsiPsi}
\psi(x,t,\zeta) = \begin{cases} \bigl([\Psi_2]_1, [\Psi_1]_2 \bigr)e^{-i\theta \sigma_3}, \qquad \text{Im}\, \zeta^2 \geq 0, \\
 \bigl([\Psi_1]_1, [\Psi_2]_2 \bigr)e^{-i\theta \sigma_3}, \qquad \text{Im}\, \zeta^2 < 0,
\end{cases}
\end{equation}
where $\bigl([\Psi_2]_1, [\Psi_1]_2 \bigr)$ denotes the matrix consisting of the first column of $\Psi_2$ together with the second column of $\Psi_1$ etc. Then $\psi$ admits the symmetries in (\ref{psisymmetry}). Indeed, uniqueness of solution of (\ref{psilax}) implies that there exist matrices $S_1(\zeta)$ and $S_2(\zeta)$ independent of $x,t$ such that
$$\sigma_3\psi(x,t,-\zeta)\sigma_3 = \psi(x,t,\zeta)S_1(\zeta),\qquad  \psi^{-1}(x,t,\zeta) = S_2(\zeta)\psi^\dagger(x,t,\bar{\zeta}),$$	
and evaluation of these equations as $x \to \pm \infty$ using (\ref{Psi12def}) shows that $S_1 = S_2 = I$. 

As a function of $\zeta$, the eigenfunction $\psi$ defined by (\ref{psiPsiPsi}) is singular for $\zeta \in \{0,Ê\infty\}$ and has a jump across the contour $\R \cup i\R$ of the form $\psi_- = \psi_+ J(\zeta)$, where $\psi_\pm$ denote the values of $\psi$ on the left and right sides of the contour and $J(\zeta)$ is a $2 \times 2$ jump matrix independent of $x$ and $t$. However, these singularities are inconsequential for the arguments below (which use the combinations $\psi_x\psi^{-1}$ and $\psi_t\psi^{-1}$). In fact, Liouville's theorem implies that $\psi$, in general, cannot be analytic over the whole Riemann $\zeta$-sphere.

\section{The dressing method}\label{dressingsec}\nequation
Starting from a seed solution $u_0$ of (\ref{GNLSgauge}) and a corresponding eigenfunction $\psi_0$ obeying the symmetries (\ref{psisymmetry}),\footnote{As outlined in the previous section, such a $\psi_0$ can be constructed from $u_0$ using only linear operations.} we seek a `dressed' eigenfunction $\psi$ of the form
\begin{equation}\label{dressing}  
  \psi = G\psi_0,
\end{equation}
where the $2\times2$-matrix valued function $G$ has the form
\begin{equation}\label{Gexpression}
  G(x,t,\zeta) = I + \sum_{j = 1}^N\left[ \frac{A_j(x,t)}{\zeta - \zeta_j} - \frac{\sigma_3 A_j (x,t) \sigma_3}{\zeta + \zeta_j}\right].
\end{equation}  
Here $\{\zeta_j, -\zeta_j\}_1^N \subset \hat{\C} \setminus \{0, \infty\}$ is a collection of simple poles with corresponding residues $\{A_j(x,t), -\sigma_3A_j(x,t)\sigma_3\}_1^N$.
The form (\ref{Gexpression}) of $G$ is motivated by the condition that $G \to I$ as $\zeta \to \infty$ and by the symmetry
\begin{equation}\label{Gsymmetry}  
  \sigma_3G(x,t,-\zeta) \sigma_3 = G(x,t,\zeta),
\end{equation}
which ascertains that the first symmetry in (\ref{psisymmetry}) is preserved by (\ref{dressing}). Moreover, in order for the second symmetry in (\ref{psisymmetry}) to be preserved by (\ref{dressing}), we require that 
\begin{equation}\label{Ginvsymmetry}
  G^{-1}(x,t,\zeta) = G^\dagger(x,t,\bar{\zeta}),
\end{equation}
i.e.
\begin{equation}\label{Ginvexpression}
  G^{-1}(x,t,\zeta) = I +  \sum_{j = 1}^N\left[ \frac{A_j^\dagger(x,t)}{\zeta - \bar{\zeta}_j} - \frac{\sigma_3 A_j^\dagger (x,t) \sigma_3}{\zeta + \bar{\zeta}_j}\right].
\end{equation}  
It follows from (\ref{Gexpression}) and (\ref{Ginvexpression}) that $G$ and $G^{-1}$ are analytic at both $\zeta = 0$ and $\zeta = \infty$.

The goal now is to show that $\psi$ satisfies (\ref{psilax}) for some matrix $U$ of the formÊ (\ref{Udef}). The compatibility of the Lax pair equations will then show that $u := U_{12}$ is a solution of equation (\ref{GNLSgauge}).

Differentiation of (\ref{dressing}) with respect to $x$ and $t$ followed by multiplication by $\psi^{-1} = \psi_0^{-1}G^{-1}$ leads to
\begin{subequations} \label{dpsipsiinv}
\begin{equation}\label{psixpsiinv}
\psi_x\psi^{-1} = G_xG^{-1} + G\psi_{0x}\psi_0^{-1}G^{-1}
\end{equation}  
and
\begin{equation}\label{psitpsiinv}
  \psi_t\psi^{-1} = G_tG^{-1} + G\psi_{0t}\psi_0^{-1}G^{-1},
\end{equation}  
\end{subequations}
respectively.

\subsection{Analysis at $\zeta = 0$}
We replace $\psi_{0x}\psi_0^{-1}$ and $\psi_{0t}\psi_0^{-1}$ in (\ref{dpsipsiinv}) with the following expressions which follow from the Lax pair equations satisfied by $\psi_0$:
\begin{equation}\label{psi0lax}
\begin{cases}
	& \psi_{0x}\psi_0^{-1} = - i\frac{1}{\zeta^2} \sigma_3 + \frac{1}{\zeta} U_{0x},	\\
	& \psi_{0t}\psi_0^{-1} = - i\left(\frac{1}{\zeta} - \frac{\zeta}{2}\right)^2 \sigma_3 + \frac{1}{\zeta} U_{0x} - \frac{i}{2}\sigma_3U_0^2 +\frac{i\zeta}{2}\sigma_3U_0.
\end{cases}
\end{equation}
Substituting into the resulting equations the expansions
$$G = G_0 + G_1 \zeta + O(\zeta^2), \quad G^{-1} = G_0^{-1} - G_0^{-1} G_1G_0^{-1}\zeta + O(\zeta^2), \qquad \zeta \to 0,$$
where $G_{0}(x,t)$ and $G_{1}(x,t)$ are independent of $\zeta$, it follows that $\psi_x\psi^{-1}$ and $\psi_t\psi^{-1}$ have double poles at $\zeta = 0$; identification of terms of $O(1/\zeta^2)$ and $O(1/\zeta)$ yields
\begin{align} \label{QR}
\psi_x\psi^{-1} = \frac{Q_{-2}}{\zeta^2} + \frac{Q_{-1}}{\zeta} + O(1),\qquad \zeta \to 0,
	\\ \nonumber
\psi_t\psi^{-1} = \frac{Q_{-2}}{\zeta^2} + \frac{Q_{-1}}{\zeta} + O(1), \qquad \zeta \to 0,
\end{align}  
where
\begin{align}\label{Qminus12}
& Q_{-2} = -iG_0\sigma_3G_0^{-1}, 
	\\ \nonumber
& Q_{-1} = i G_0 \sigma_3 G_0^{-1} G_1G_0^{-1} - iG_1\sigma_3G_0^{-1} + G_0 U_{0x} G_0^{-1}.
\end{align}
On the other hand, from (\ref{Gexpression}) we find the following expressions for $G_0$ and $G_1$:
$$G_0 = I - \sum_{j = 1}^N\frac{A_j + \sigma_3A_j \sigma_3}{\zeta_j}, \qquad
G_1 = - \sum_{j = 1}^N\frac{A_j - \sigma_3A_j \sigma_3}{\zeta_j^2},$$
so that the matrices $G_0$ and $G_1$ are diagonal and off-diagonal, respectively. Thus the expressions for $Q_{-2}$ and $Q_{-1}$ in (\ref{Qminus12}) simplify to
\begin{equation}\label{Qminus12simple}
Q_{-2} = -i\sigma_3, \qquad 
Q_{-1} = -2iG_1G_0^{-1}\sigma_3 + G_0 U_{0x} G_0^{-1}.
\end{equation}
In particular, $Q_{-1}$ is an off-diagonal matrix. 

\subsection{Analysis at $\zeta = \infty$}
We again replace $\psi_{0x}\psi_0^{-1}$ and $\psi_{0t}\psi_0^{-1}$ in (\ref{dpsipsiinv}) with the expressions in (\ref{psi0lax}). Substituting into the resulting equations the expansions
\begin{align}\nonumber
 & G = I + \frac{G_{-1}}{\zeta} + \frac{G_{-2}}{\zeta^2} + O(1/\zeta^3), \qquad \zeta \to \infty,
	\\ \label{Ginvexp}
& G^{-1} = I - \frac{G_{-1}}{\zeta} + \frac{G_{-1}^2 - G_{-2}}{\zeta^2} +  O(1/\zeta^3), \qquad \zeta \to \infty,
\end{align} 
where $G_{-1}$ and $G_{-2}$ are independent of $\zeta$, it follows that $\psi_x\psi^{-1} \to 0$ as $\zeta \to \infty$ whereas $\psi_t\psi^{-1}$ has a double pole at $\zeta = \infty$; identification of terms of $O(\zeta^n)$, $n = 0,1,2$, yields
$$\psi_t\psi^{-1} = -\frac{i}{4}\sigma_3 \zeta^2 + Q_1\zeta + Q_0 + O(1/\zeta), \qquad \zeta \to \infty,$$
where
\begin{align}\label{Q10}
& Q_1 = \frac{i}{4}[\sigma_3, G_{-1}] + \frac{i}{2}\sigma_3 U_0,
	\\\nonumber
& Q_0 = \frac{i}{4}[\sigma_3, G_{-2}] -\frac{i}{4}\sigma_3G_{-1}^2 
+ \frac{i}{4}G_{-1}\sigma_3G_{-1}
  + \frac{i}{2}G_{-1}\sigma_3U_0 - \frac{i}{2}\sigma_3U_0G_{-1}
  	\\ \nonumber
& \qquad  \; + i\sigma_3 - \frac{i}{2}\sigma_3 U_0^2.
\end{align}
On the other hand, from (\ref{Gexpression}) we find the following expressions for $G_{-1}$ and $G_{-2}$:
\begin{align}\label{Gminus12}
 G_{-1} = \sum_{j = 1}^N\left(A_j - \sigma_3 A_j \sigma_3\right),
\qquad  G_{-2} = \sum_{j = 1}^N\zeta_j \left(A_j + \sigma_3 A_j \sigma_3\right),
\end{align}
so that the matrices $G_{-1}$ and $G_{-2}$ are off-diagonal and diagonal, respectively. Thus the expressions for $Q_1$ and $Q_0$ in (\ref{Q10}) simplify to
\begin{align}\label{Q10simple}
Q_1 = \frac{i}{2}\sigma_3 U, \qquad Q_0 = -\frac{i}{2}U^2 \sigma_3 + i\sigma_3,
\end{align}
where we have defined
\begin{equation}\label{UGU0}
  U := G_{-1} + U_0.
\end{equation}
Equations (\ref{Ginvexpression}) and (\ref{Ginvexp}) yield the following alternative expression for $G_{-1}$:
$$G_{-1} = - \sum_{j = 1}^N\left(A_j^\dagger - \sigma_3 A_j^\dagger \sigma_3\right).$$
Thus $(G_{-1})_{21} = -\overline{(G_{-1})}_{12}$, so that $U$ is of the form (\ref{Udef}) with $u := (G_{-1})_{12} + u_0$. 

\subsection{The generated solution}
It follows from the previous two subsections that the functions
\begin{subequations}\label{psiQfuns}
\begin{equation}
\psi_x\psi^{-1} - \left(\frac{Q_{-2}}{\zeta^2} + \frac{Q_{-1}}{\zeta}\right)
\end{equation}
and
\begin{equation}
\psi_t\psi^{-1} - \left(\frac{Q_{-2}}{\zeta^2} + \frac{Q_{-1}}{\zeta} + Q_0  + Q_1\zeta -\frac{i}{4}\sigma_3 \zeta^2\right)
\end{equation}
\end{subequations}
are analytic near $\zeta =0$ and $\zeta = \infty$. Thus, in view of (\ref{dpsipsiinv}) and the expressions (\ref{Gexpression}) and (\ref{Ginvexpression}) for $G$ and $G^{-1}$, they are analytic on the whole Riemann $\zeta$-sphere except possibly at points in the set $\{\pm\zeta_j, \pm\bar{\zeta}_j\}_1^N$. We claim that the $A_j$'s in (\ref{Gexpression}) can be chosen so that the functions in (\ref{psiQfuns}) are analytic everywhere. Assume that such a choice has been made. Then, since the functions  in (\ref{psiQfuns}) tend to zero as $\zeta \to \infty$, Liouville's theorem implies that they both vanish identically. Together with the expressions (\ref{Qminus12simple}) and (\ref{Q10simple}) for the coefficients $Q_{-2}, Q_0, Q_1$, this implies that $\psi$ satisfies the following pair of equations:
\begin{equation}\label{psilaxQ}
\begin{cases}
	& \psi_x + i\frac{1}{\zeta^2} \sigma_3 \psi = \frac{1}{\zeta} Q_{-1} \psi,	\\
	& \psi_t + i\left(\frac{1}{\zeta} - \frac{\zeta}{2}\right)^2 \sigma_3 \psi = \left(\frac{1}{\zeta} Q_{-1} - \frac{i}{2}\sigma_3U^2 +\frac{i\zeta}{2}\sigma_3U\right)\psi,
\end{cases}
\end{equation}
where $U$ has the form (\ref{Udef}). Using that $Q_{-1}$ is off-diagonal and identifying terms of $O(\zeta)$ in the compatibility equation $\psi_{xt} = \psi_{tx}$, we find that
\begin{equation}\label{QUx}  
  Q_{-1} = U_x.
\end{equation}
Consequently, (\ref{psilaxQ}) has exactly the form of the Lax pair (\ref{psilax}) and its compatibility implies that the function $u = U_{12}$ satisfies (\ref{GNLSgauge}). We infer from (\ref{Gminus12}) and (\ref{UGU0}) that the generated solution $u(x,t)$ is given by
\begin{equation}\label{uAju0}
u(x,t) = 2\sum_{j = 1}^N(A_j(x,t))_{12} + u_0(x,t).
\end{equation}

\subsection{The dressing transformation}
In order to complete the construction of the map $u_0 \mapsto u$, it only remains to determine the structure of the $A_j$'s in (\ref{uAju0}) so that the functions $\psi_x\psi^{-1}$ and $\psi_t\psi^{-1}$ are regular at the points $\{\pm \zeta_j, \pm \bar{\zeta}_j\}_1^N$. To this end we consider the following series of $N$ consecutive dressing transformations, each of which adds two poles:
\begin{equation}\label{GDDD}
G =  D_ND_{N-1}\cdots D_1, \qquad \psi_j = D_j\psi_{j-1}, \quad j = 1, \dots, N,
\end{equation}
where, for $j = 1, \dots, N$,
\begin{subequations}\label{Djdef}
\begin{align} \label{Djdef1}
& D_j(x,t,\zeta) = I + \frac{B_j(x,t)}{\zeta - \zeta_j} - \frac{\sigma_3 B_j(x,t) \sigma_3}{\zeta + \zeta_j},
	\\ \label{Djdef2}
& D_j^{-1}(x,t,\zeta) = I + \frac{B_j^\dagger(x,t)}{\zeta - \bar{\zeta}_j} - \frac{\sigma_3 B_j^\dagger(x,t) \sigma_3}{\zeta + \bar{\zeta}_j},
\end{align}
\end{subequations}
and $\{B_j(x,t)\}_1^N$ are a set of $2\times2$-matrix valued functions independent of $\zeta$.

\begin{lemma}\label{Bjlemma}
Let $\{b_j\}_1^N$ be a set of $N$ nonzero complex constants.
Define the functions $B_j(x,t)$, $j = 1,\dots, N$, inductively by
\begin{equation}\label{eq26}
  B_j(x,t) = | z_j(x,t) \rangle \langle y_j(x,t) |,
\end{equation}  
where the column vector $ | z_j \rangle = \langle z_j |^\dagger$ and the row vector $\langle y_j | = | y_j \rangle^\dagger$ are defined in terms of the $(j - 1)$th eigenfunction $\psi_{j-1}$ by\footnote{Here and in some equations below the $(x,t)$-dependence is suppressed.} 
\begin{align} \label{eq35}
 & \langle y_j |  = \begin{pmatrix} b_j & b_j^{-1} \end{pmatrix} \psi_{j - 1}^{-1}(\zeta_j),
	\\ \label{gjhjprime}
 & | z_j \rangle = \frac{\zeta_j^2 - \bar{\zeta}_j^2}{2} \begin{pmatrix} \alpha_j & 0 \\ 0 & \bar{\alpha}_j \end{pmatrix} | y_j \rangle, \quad \text{where} \quad \alpha_j^{-1} = \langle y_j | \begin{pmatrix} \zeta_j & 0 \\ 0 & \bar{\zeta}_j \end{pmatrix} |y_j\rangle.
\end{align}
Define $\psi := \psi_N$ by (\ref{GDDD})-(\ref{gjhjprime}). Then the functions $\psi_x\psi^{-1}$ and $\psi_t\psi^{-1}$ are analytic at the points in the set $\{\pm \zeta_j, \pm \bar{\zeta}_j\}_1^N$.
\end{lemma}
\proofbegin
The proof proceeds by induction. Suppose that $\psi_{j-1}$ has been defined by (\ref{GDDD})-(\ref{gjhjprime}) and that the functions $\psi_{j-1,x}\psi_ {j-1} ^{-1}$ and $\psi_{j-1,t}\psi_{j-1}^{-1}$ are analytic at $\zeta = \pm \zeta_i$ and $\zeta = \pm \bar{\zeta}_i$ for $i = 1, \dots, j-1$. We will show that $\psi_{jx}\psi_j^{-1}$ and $\psi_{jt}\psi_j^{-1}$, where $\psi_j = D_j \psi_{j-1}$,Ê are analytic at $\zeta = \pm \zeta_i$ and $\zeta = \pm \bar{\zeta}_i$ for $i = 1, \dots, j$.

Differentiation of $\psi_j = D_j \psi_{j-1}$ yields
\begin{equation}\label{psijxpsijinv}
  \psi_{jx}\psi_j^{-1} = D_{jx}D_j^{-1} + D_j\psi_{j-1,x}\psi_{j-1}^{-1}D_j^{-1}.
\end{equation}
The assumption on $\psi_{j-1}$ together with (\ref{Djdef}) implies that the right-hand side is analytic at the points 
$\{\pm \zeta_i, \pm \bar{\zeta}_i\}_1^{j-1}$. To show that $\psi_{jx}\psi_j^{-1}$ is regular at $\zeta_j$, we note that (\ref{psijxpsijinv}) implies that $\psi_{jx}\psi_j^{-1}$ has at most a simple pole at $\zeta_j$, and that
\begin{equation}\label{reszetaj}  
  \underset{\zeta_j}{\text{Res}}\left(\psi_{jx}\psi_j^{-1}\right) = (B_j \psi_{j-1}(\zeta_j))_x\psi_{j-1}^{-1}(\zeta_j)D_j^{-1}(\zeta_j).
\end{equation}
Using (\ref{eq26}) and (\ref{eq35}), we can write (\ref{reszetaj}) as
$$\underset{\zeta_j}{\text{Res}}\left(\psi_{jx}\psi_j^{-1}\right) = |z_j\rangle_x \langle y_j | D_j^{-1}(\zeta_j).$$
We claim that 
\begin{equation}\label{yjDjinv0}
  \langle y_j | D_j^{-1}(\zeta_j) =0, 
\end{equation}  
so that the residue vanishes. Indeed, by (\ref{Djdef2}), 
\begin{equation}\label{yangleeq}
\langle y_j | D_j^{-1}(\zeta_j)  = \langle y_j | + \frac{\langle y_j | y_j \rangle}{\zeta_j - \bar{\zeta}_j} \langle z_j | - \frac{\langle y_j | \sigma_3 | y_j\rangle}{\zeta_j + \bar{\zeta}_j} \langle z_j | \sigma_3,
\end{equation}
and the right-hand side of this equation vanishes by virtue of (\ref{gjhjprime}).
Since the symmetry properties (\ref{psisymmetry}) are preserved by (\ref{GDDD}), we deduce that $\psi_{jx}\psi_j^{-1}$ is analytic also at $-\zeta_j$ and $\pm \bar{\zeta}_j$. Similar arguments establish the regularity of $\psi_{jt}\psi_j^{-1}$.
\proofend

By (\ref{Gexpression}) and (\ref{GDDD}),
\begin{equation}\label{eq38}
  A_j = \underset{\zeta_j}{\text{Res}}\, G(\zeta)
= D_N(\zeta_j) \dots D_{j+1}(\zeta_j)B_jD_{j-1}(\zeta_j) \dots D_1(\zeta_j).
\end{equation}
On the other hand, it follows from Lemma \ref{Bjlemma} that 
\begin{equation}\label{sigmaBjsigma}
  \sigma_2 B_j^T(x,t) \sigma_2 = \frac{\zeta_j^2 - \bar{\zeta}_j^2}{2\zeta_j} D_j^{-1}(x,t,\zeta_j), \qquad \sigma_2 = \begin{pmatrix} 0	&	-i \\ i	& 0\end{pmatrix},
\end{equation}  
and
\begin{equation}\label{sigmaDjsigma}
  \sigma_2 D_j(x,t,\zeta)^T \sigma_2 = \frac{\zeta^2 - \bar{\zeta}_j^2}{\zeta^2 - \zeta_j^2} D_j^{-1}(x,t,\zeta).
\end{equation}
Using (\ref{sigmaBjsigma}) and (\ref{sigmaDjsigma}), the expression (\ref{eq38}) for $A_j$ can be written as
\begin{equation}\label{eq39}  
  A_j(x,t) = \frac{1}{a_j}\sigma_2 G^{-1}(x,t,\zeta_j)^T \sigma_2
\end{equation}
where the complex constants $\{a_j\}_1^N$ are defined by
\begin{equation}\label{ajdef}
a_j = \frac{2\zeta_j}{\zeta_j^2 - \bar{\zeta}_j^2} \prod_{k \neq j}\frac{\zeta_j^2 - \zeta_k^2}{\zeta_j^2 - \bar{\zeta}_k^2}.
\end{equation}
Substitution of (\ref{eq39}) into the expression (\ref{Gexpression}) for $G$ yields
\begin{equation}\label{eq41}  
  G(x,t,\zeta) = I + \sum_{j = 1}^N \frac{1}{a_j} \left[ \frac{\sigma_2 G^{-1}(x,t,\zeta_j)^T \sigma_2}{\zeta - \zeta_j} - \frac{\sigma_3 \sigma_2 G^{-1}(x,t,\zeta_j)^T \sigma_2 \sigma_3}{\zeta + \zeta_j}\right],
\end{equation}
Eliminating $A_j$ from (\ref{eq38}) and (\ref{eq39}), and using Lemma \ref{Bjlemma} together with the fact that $\psi_{j-1}^{-1}D_{j-1} \cdots D_1 = \psi_0^{-1}$ in the resulting equation, we find that 
\begin{equation}\label{eq42}  
  \sigma_2G^{-1}(x,t,\zeta_j)^T\sigma_2 = \begin{pmatrix} r_j(x,t) \\ s_j(x,t) \end{pmatrix}\begin{pmatrix} b_j & b_j^{-1} \end{pmatrix} \psi_0^{-1}(x,t,\zeta_j)
\end{equation}
for some functions $r_j(x,t), s_j(x,t)$, $j = 1, \dots, N$.
Evaluating equation (\ref{eq41}) at $\zeta = \bar{\zeta}_k$, and using (\ref{eq42}) we find
\begin{align}\label{eq44}  
  G(x,t,\bar{\zeta}_k) = I + \sum_{j = 1}^N \frac{1}{a_j} \biggl[& \frac{1}{\bar{\zeta}_k - \zeta_j} 
  \begin{pmatrix} r_j(x,t) \\ s_j(x,t) \end{pmatrix}\begin{pmatrix} b_j & b_j^{-1} \end{pmatrix} \psi_0^{-1}(x,t,\zeta_j) 
	\\ \nonumber
& - \frac{1}{\bar{\zeta}_k + \zeta_j} \sigma_3 \begin{pmatrix} r_j(x,t) \\ s_j(x,t) \end{pmatrix}\begin{pmatrix} b_j & b_j^{-1} \end{pmatrix} \psi_0^{-1}(x,t,\zeta_j) \sigma_3 \biggr].
\end{align}
But the symmetry (\ref{Ginvsymmetry}) together with (\ref{eq42}) shows that
\begin{equation}\label{eq43}  
  G(x,t,\bar{\zeta}_k) = \sigma_2 \begin{pmatrix} \overline{r_k(x,t)} \\ \overline{s_k(x,t)} \end{pmatrix}\begin{pmatrix} \bar{b}_k & \bar{b}_k^{-1} \end{pmatrix} \overline{\psi_0^{-1}(x,t,\zeta_k)}\sigma_2,
\end{equation}
so that multiplying equation (\ref{eq44}) by $\sigma_2\overline{\psi_0(x,t,\zeta_k)} \begin{pmatrix} \bar{b}_k^{-1} & -\bar{b}_k\end{pmatrix}^T$ from the right, the left-hand side disappears and we are left with
\begin{align}\label{eq46}  
  0 =& \sigma_2\overline{\psi_0(x,t,\zeta_k)} \begin{pmatrix} \bar{b}_k^{-1} \\ -\bar{b}_k\end{pmatrix}
  + \sum_{j = 1}^N \frac{1}{a_j} \biggl[ \frac{1}{\bar{\zeta}_k - \zeta_j} 
  \begin{pmatrix} r_j(x,t) \\ s_j(x,t) \end{pmatrix}\begin{pmatrix} b_j & b_j^{-1} \end{pmatrix} \psi_0^{-1}(x,t,\zeta_j) 
  	\\ \nonumber
& - \frac{1}{\bar{\zeta}_k + \zeta_j} \sigma_3 \begin{pmatrix} r_j(x,t) \\ s_j(x,t) \end{pmatrix}\begin{pmatrix} b_j & b_j^{-1} \end{pmatrix} \psi_0^{-1}(x,t,\zeta_j) \sigma_3 \biggr]\sigma_2\overline{\psi_0(x,t,\zeta_k)} \begin{pmatrix} \bar{b}_k^{-1} \\ -\bar{b}_k\end{pmatrix}.
\end{align}
These are $2N$ algebraic scalar equations for the $2N$ unknowns $\{r_j(x,t), s_j(x,t)\}_1^N$. Given the seed eigenfunction $\psi_0$, the solution of (\ref{eq46}) expresses the $r_j$'s and the $s_j$'s in terms of $\{\zeta_j, b_j\}_1^N$. Thus, by (\ref{eq39}) and (\ref{eq42}), the $A_j$'s are determined in terms of the $2N$ complex constants $\{\zeta_j, b_j\}_1^N$. 
This yields the following result.

\begin{proposition}
Let $N$ be a positive integer and let $\{\zeta_j, b_j\}_{j = 1}^N$ be nonzero complex constants such that $\zeta_j \neq \zeta_k$ for $j \neq k$.
Assume that $u_0(x,t)$ satisfies equation (\ref{GNLSgauge}) and let $\psi_0(x,t,\zeta)$ be an associated eigenfunction obeying the symmetries (\ref{psisymmetry}).
Then the following function $u(x,t)$ is also a solution of equation (\ref{GNLSgauge}):
\begin{equation}\label{ufinal}  
  u(x,t) = 2\sum_{j = 1}^N(A_j(x,t))_{12} + u_0(x,t),
\end{equation}
where the $2\times2$-matrix valued function $A_j(x,t)$ is given by
\begin{equation}\label{Ajfinal}
  A_j(x,t) = \frac{1}{a_j} \begin{pmatrix} r_j(x,t) \\ s_j(x,t) \end{pmatrix}\begin{pmatrix} b_j & b_j^{-1} \end{pmatrix} \psi_0^{-1}(x,t,\zeta_j),
\end{equation}
the constants $\{a_j\}_1^N$ are given by (\ref{ajdef}), and $\{r_j(x,t), s_j(x,t)\}_{j = 1}^N$ are determined by the linear algebraic system of equations (\ref{eq46}).
\end{proposition}

\section{Solitons}\label{solitonsec}\nequation
In this section we apply the dressing method described in the previous section to the particular seed solution $u_0 = 0$ with corresponding eigenfunction $\psi_0 = \exp(-i\theta \sigma_3)$. The addition of $2N$ poles to $\psi_0$ yields the $N$-soliton solution of (\ref{GNLSgauge}). Since in this case the solution of the algebraic system (\ref{eq46}) can be expressed in a simple form, we find simple explicit formulas for the solitons.

Recall that $\theta = \theta(x,t,\zeta)$ was defined in (\ref{thetadef}). Substituting $u_0 = 0$ and $\psi_0 = \exp(-i\theta \sigma_3)$ into the equations (\ref{ufinal}) and (\ref{Ajfinal}), we find
\begin{equation}\label{usol1}
  u(x,t) = 2\sum_{j = 1}^N\frac{r_j}{a_j f_j},
\end{equation}
where $f_j := f_j(x,t) = b_j e^{i\theta(x,t,\zeta_j)}$, $j = 1, \dots, N$.
On the other hand, for this choice of $\psi_0$, the top entry of equation (\ref{eq46}) yields
\begin{equation}\label{eq66}
  2\sum_{j=1}^N \frac{r_j K_{jk}}{a_j f_j} = \bar{f}_k^2,\qquad k = 1, \dots, N,
\end{equation}  
where the entries of the $N\times N$-matrix $K = K(x,t)$ are defined by
\begin{equation}\label{Kdef}
  K_{jk} = \frac{1}{\zeta_j^2 - \bar{\zeta}_k^2} (\zeta_j f_j^2 \bar{f}_k^2 + \bar{\zeta}_k), \qquad j,k = 1, \dots, N.
\end{equation}  
Combining (\ref{usol1}) and (\ref{eq66}), we find the following expression for the $N$-soliton solution.

\begin{proposition}\label{solprop}
The $N$-soliton solution $u_N(x,t)$ of the generalized NLS equation (\ref{GNLSgauge}) is given explicitly by
\begin{equation}\label{uNfinal}
  u_N(x,t) = \sum_{k,j = 1}^N \bar{f}_k^2(K^{-1})_{kj},
\end{equation}
where the $N\times N$-matrix $K = K(x,t)$ is defined by (\ref{Kdef}), the coefficients $\{f_j\}_1^N$ are given by
\begin{equation}\label{fjinprop}
f_j := f_j(x,t) = b_j \exp\left(i \frac{1}{\zeta_j^2}x + i\left(\frac{1}{\zeta_j} - \frac{\zeta_j}{2}\right)^2t\right), \qquad j = 1, \dots, N,
\end{equation}
and the solution depends on the $2N$ complex parameters $\{b_j, \zeta_j\}_1^N$.
\end{proposition}

\section{DNLS solitons}\label{DNLSsolitonsec}\nequation
The derivative nonlinear Schr\"odinger equation (\ref{DNLS}) admits the Lax pair \cite{K-N}
\begin{equation}\label{DNLSlax}
\begin{cases}
 & \psi_x + i \frac{1}{\zeta^2} \sigma_3\psi = \frac{1}{\zeta}Q\psi, 	
  	\\
 &  \psi_t + 2 i \frac{1}{\zeta^4}\sigma_3 \psi = \left(\frac{2}{\zeta^3}Q - \frac{i}{\zeta^2}Q^2 \sigma_3 - \frac{1}{\zeta}(iQ_x \sigma_3 - Q^3)\right)\psi,
\end{cases}
\end{equation}
where the $2\times 2$-matrix valued function $Q(x,t)$ is given by
\begin{equation}\label{Qdef}
  Q = \begin{pmatrix} 0 & q \\ -\bar{q} & 0 \end{pmatrix}.
\end{equation}
If we identify $Q$ with $U_x$, then the $x$-part of this Lax pair coincides with the $x$-part of the Lax pair (\ref{psilax}). 
A similar analysis that led to expression (\ref{uNfinal}) for $u_N$ applied to equations (\ref{Qminus12simple}) and (\ref{QUx}) yields (cf. \cite{HC})
\begin{equation}\label{uNxfinal}
  u_{Nx} = -2i\left(\sum_{k,j = 1}^N \frac{\bar{f}_k^2}{\zeta_j^2} (K^{-1})_{kj}\right)\left(1 + \sum_{k,j = 1}^N \frac{1}{\bar{\zeta}_k} (K^{-1})_{kj}\right),
\end{equation}
where $K$ and the $f_j$'s are defined as in Proposition \ref{solprop}. In view of the identification of $Q$ with $U_x$, it is natural to expect the $N$-soliton solution $q_N$ of the DNLS equation to be given by an expression similar to the right-hand side of (\ref{uNxfinal}). In fact, it was shown in \cite{HC} that $q_N(x,t)$ is given by the right-hand side of (\ref{uNxfinal}) with $f_j(x,t)$ replaced by
\begin{equation}\label{fjDNLS}
  f_j(x,t) = b_je^{i(\zeta_j^{-2}x + 2\zeta_j^{-4}t)}.
\end{equation}
Now equations (\ref{uNfinal}) and (\ref{uNxfinal}) imply that the right-hand side of (\ref{uNxfinal}) is the $x$-derivative of the right-hand side of (\ref{uNfinal}) when $f_j$ is given by (\ref{fjinprop}). Since all the time-dependence in these equations lies in the exponents of $f_j$ and $\bar{f}_j$ and is irrelevant for the differentiation with respect to $x$, this property holds also when the $f_j$'s are given by (\ref{fjDNLS}). This yields the following simple expressions for the DNLS $N$-solitons.

\begin{proposition}
The $N$-soliton solution $q_N(x,t)$ of the DNLS equation (\ref{DNLS}) is given explicitly by
\begin{equation}\label{DNLSsoliton}  
  q_N(x,t) = \frac{\partial}{\partial x}\left(\sum_{k, j = 1}^N \bar{f}_k^2 (K^{-1})_{kj}\right),
\end{equation}
where the $N\times N$-matrix $K = K(x,t)$ is defined by (\ref{Kdef}), the coefficients $\{f_j\}_1^N$ are defined by (\ref{fjDNLS}), and the solution depends on the $2N$ complex parameters $\{b_j, \zeta_j\}_1^N$.
\end{proposition}

It can be checked explicitly for small values of $N$ that the solitons $u_N$ and $q_N$ given by (\ref{uNfinal}) and (\ref{DNLSsoliton}) indeed satisfy the generalized NLS equation (\ref{GNLSgauge}) and the DNLS equation (\ref{DNLS}), respectively.

\section{Conclusions}
We have implemented the dressing method to equation (\ref{GNLSgauge}), which is equivalent to the generalized NLS equation (\ref{GNLS}). This provides a way of generating new solutions from already known ones. In particular, starting with the trivial solution $u_0 \equiv 0$, we arrived at the explicit expression (\ref{uNfinal}) for the $N$-soliton solution. Since (\ref{GNLS}) is related by a gauge transformation to a member of the DNLS hierarchy, the construction bears similarities with the implementation of the dressing method to DNLS \cite{HC} (see also \cite{X}). The presence of additional singularities in the $t$-part of the Lax pair makes the argument for (\ref{GNLSgauge}) somewhat more complicated.

The link between equations (\ref{GNLS}) and (\ref{DNLS}) allowed us to find a simplification (see equation (\ref{DNLSsoliton})) of the formulas for the DNLS $N$-solitons presented in \cite{HC}. The fact that the right-hand side of (\ref{DNLSsoliton}) can be expressed as a partial $x$-derivative is related to the fact that the DNLS hierarchy can be reformulated in terms of the potential $u = \partial_x^{-1}q$. Consequently, the solitons of members of this hierarchy can be expressed as $x$-derivatives of elementary functions cf. \cite{Tsu}.

\bigskip
\noindent
{\bf Acknowledgement} {\it The author thanks the referees for valuable suggestions, A. S. Fokas for helpful discussions, and T. Tsuchida for bringing the reference \cite{Tsu} to his attention.}

\bibliography{is}

\end{document}